\documentclass[useAMS,usenatbib]{mnras}

\usepackage{amsmath,amssymb,mathrsfs,graphicx,float,latexsym,url}
\usepackage{epstopdf,ragged2e}
\usepackage{natbib,url}
\usepackage{hyperref}

\usepackage[usenames,dvipsnames]{color}
\definecolor{darkred}{rgb}{0.5,0,0}
\definecolor{darkgreen}{rgb}{0,0.5,0}
\definecolor{darkblue}{rgb}{0,0,0.5}
\definecolor{prussian}{rgb}{0.0, 0.19, 0.33}
\definecolor{richelectricblue}{rgb}{0.03, 0.57, 0.82}
\definecolor{teal}{rgb}{0.0, 0.5, 0.5}
\definecolor{mediumseagreen}{rgb}{0.24, 0.7, 0.44}
\definecolor{lust}{rgb}{0.9, 0.13, 0.13}
\definecolor{ballblue}{rgb}{0.13, 0.67, 0.8}
\definecolor{darkcyan}{rgb}{0.0, 0.55, 0.55}
\definecolor{mountainmeadow}{rgb}{0.19, 0.73, 0.56}
\definecolor{palecarmine}{rgb}{0.69, 0.25, 0.21}
\definecolor{richcarmine}{rgb}{0.84, 0.0, 0.25}
\definecolor{tangelo}{rgb}{0.98, 0.3, 0.0}
\definecolor{venetian}{rgb}{0.784,0.031,0.082}
\definecolor{bdfrance}{rgb}{0.192,0.549,0.906}
\usepackage{mathtools}
\usepackage{amsbsy}
\usepackage{bm}
\usepackage{float}
\newcommand{\nus}{\nu_{\star}}
\newcommand{\bj}{\boldsymbol{J}}
\newcommand{\Bs}{B_{\star}}
\newcommand{\Rs}{R_{\star}}
\newcommand{\Ms}{M_{\star}}

\newcommand{\bb}{\boldsymbol{B}}
\newcommand{\bdb}{\delta \boldsymbol{B}}

\def\apj{{ApJ}}

\def\apjl{{ApJL}}

\def\aap{{A\&A}}

\def\mnras{{MNRAS}}

\def\pasj{{Publications of the Astronomical Society of Japan}}

\def\nat{{Nature}}

\def\prl{{Phys. Rev. Lett.}}

\def\04a{{2004 a}}
\def\04b{{2004 b}}

\begin{document}

%Outbursts in 1E 1547-5408 via magnetic shifts
%\title[Outburst event in 1E 1547-5408 via magnetic shifts]
\title[Outbursts in 1E 1547-5408 via magnetic shifts]
{The radio shut-off, glitch, and X-ray burst in 1E 1547.0$-$5408 interpreted through magnetic reconfiguration}
\author[A. G.~Suvorov]{Arthur G. Suvorov$^{1,2}$\thanks{arthur.suvorov@manlyastrophysics.org}\\
$^1$Manly Astrophysics, 15/41-42 East Esplanade, Manly, NSW 2095, Australia\\
$^2$Theoretical Astrophysics, Eberhard Karls University of T{\"u}bingen, T{\"u}bingen, D-72076, Germany
}

\date{Accepted ?. Received ?; in original form ?}

\pagerange{\pageref{firstpage}--\pageref{lastpage}} \pubyear{?}

\maketitle
\label{firstpage}

\begin{abstract}

\noindent A short X-ray burst was observed from the radio-loud magnetar 1E 1547.0$-$5408 in 2022 April. Unusually however, the source stopped showing radio pulsations $\gtrsim 3\,$weeks \emph{prior} to the burst. After recovery, radio timing revealed that the object had also undergone a modest glitch. A model for the overall event is constructed where an initially mild perturbation adjusts the magnetic geometry near the polar caps, leading to shallow fractures. Crustal ejecta or particles leaking from a pair-plasma fireball pollute the magnetospheric gaps, shutting off the pulsar mechanism, but the energy release is not yet large enough to noticeably enhance the X-ray flux. This perturbation gradually ramps, eventuating in a large-scale energy redistribution which fuels the burst. The star's mass quadrupole moment changes in tandem, issuing a glitch. Some quantitative estimates for the magnetic reconfiguration under this interpretation are provided, based on a quasi-static model where the fluid evolves through a sequence of hydromagnetic equilibria. 

\end{abstract}


%%%%%%%%%%%%%%%%%%%%%%%%%%%%%%%%%

\section{Introduction} 
\label{sec:intro}

The magnetar class of neutron stars are defined by their propensity for burst and flare activity in the X- and gamma-ray bands. As put forth by \cite{td92} and \cite{td93}, this activity is likely powered by a decaying, ultra-strong ($\gtrsim 10^{15}\,$G) magnetic field. Several Galactic magnetars also transiently operate as radio pulsars, providing valuable information about their magnetospheric structure \citep{med10,low21}. In the high-$B$ pulsar PSR J1119--6127, radio shut-off was observed following high-energy X-ray activity in 2016, with the source then reactivating $\sim$~2 weeks later \citep{arch17,dai18}. The cessation of radio activity was interpreted by \cite{arch17} to result from particles in the trapped pair-plasma `fireball', fuelled by dissipated magnetic energy, straying into the magnetospheric gaps and screening the acceleration potential \citep{td95}.

\cite{low23} recently reported that the magnetar 1E 1547.0--5408 (henceforth, E1547) became radio-quiet some weeks \emph{prior} to emitting a burst in April 2022. This is indicative that inconspicuous magnetic activity takes place before magnetar outbursts. The (unabsorbed) X-ray flux then spiked from $\approx 1.7 \times 10^{-11} \text{ erg s}^{-1} \text{cm}^{-2}$, which the source has persistently displayed since its 2009 outburst \citep{coti20}, to $\approx 6.0 \times 10^{-11} \text{ erg s}^{-1} \text{cm}^{-2}$ for $\sim 17$~days. Once the source stabilised, returning to its circa 2009 flux level, and began pulsating in the radio band once more, its spin frequency had jumped by $\approx 0.2 \, \mu$Hz relative to its well measured previous value. Approximately $\sim 6\%$ of the observed pulsar population, including many magnetars \citep{fuen17}, undergo such glitch events: sudden increases in spin frequency, typically followed by an episode of accelerated spin-down \citep{hm15,zhou22}. %\cite[see][for a review]{zhou22}. %PSR J1119--6127 also exhibited a large glitch after its flare in 2016 \citep{arch16}.

While glitch activity in `ordinary' pulsars is thought to be associated with the unpinning of superfluid vortices \cite[e.g.,][]{hm15}, magnetic reconfigurations may be responsible for at least some magnetar (anti-)glitches \citep{i01,grs15,msm15b}. Magnetic energy redistributions may non-negligibly shift the stellar moment of inertia, thereby adjusting the rotation rate by angular momentum conservation. Moreover, Maxwell stresses associated with this reconfiguration may be strong enough to break (or instigate some kind of failure) within the crystalline crust in a spatially dependent way that depends on the field particulars. Tectonic activity is often invoked in explanations of bursts and flares from magnetars generally \cite[e.g.,][]{lan15,suvk19}.

In this short paper, we present a simple, analytic model of a growing magnetic perturbation to interpret the sequence of events seen in E1547. An initially mild reconfiguration results in shallow fractures that are confined to the polar caps, halting radio activity once gap zones are polluted. Importantly, the fractures are sufficiently minor such that energy releases are not high enough to be discernible against noise in the already high X-ray flux from the source. We stipulate that avalanche-like behaviour then results in an intense reconfiguration some time later \cite[via Hall-wave propagation, as in the model of][]{xi16}, unleashing an X-ray burst from a larger quake while adjusting the stellar ellipticity, inducing a glitch and accelerating spin-down. 

This work is organised as follows. {Section \ref{sec:hall} provides an overview of the theoretical mechanisms considered here.} In Section \ref{sec:hydrostruc} we introduce magnetohydrodynamic (MHD) preliminaries, necessary to calculate astrophysical observables associated with E1547.  Section \ref{sec:outburst} discusses key aspects of the sequential observations and their relationship to the magnetar's properties. We assume canonical values for the stellar mass, $M = 1.4 M_{\odot}$ ($M_{1.4}=1$), and radius, $\Rs = 10^{6}$~cm ($R_{6}=1$), throughout.

\section{Theoretical overview and ramping perturbations}
\label{sec:hall}

{As the intense magnetic field gradually evolves, mechanical stresses are built up in the elastic crust of a magnetar. It is thought that localised regions eventually become overstrained, resulting in failures that release magnetic and elastic energy \cite[e.g.,][]{lan15}. Stress is subsequently relieved from the failed zone as a plastic flow is initiated \citep{hk09}. Importantly, the release of magnetic energy implies a decrease in the local $B$-field strength. As pointed out by \cite{xi16}, this rapid adjustment generates currents that can drag the field lines and excite Hall waves (i.e. propagating solutions to the non-ideal induction equation). These authors showed that the peak amplitude of Hall waves typically exceeds that of the ambient field (see Figure 3 therein), and thus wave crests can drive a sequence of overstraining events whereupon new Hall waves are excited. The propagation of Hall waves, together with the thermoplastic waves roused by the failures themselves \citep{bell14}, thus instigate chains of failure within the crust.}

{Introducing the Hall, $D_{\rm H} = c B / \left(4 \pi e n_{e}\right)$, and thermal, $D_{\rm T} = \kappa / C_{V}$, diffusion coefficients, \cite{xi16} deduce that the two relevant speeds are $v_{\rm H} \sim (\alpha_{r} D_{\rm H})^{1/2} \approx 2 \times 10^{-2} \left(\alpha_{r,-2} B_{14}/n_{e,34}\right)^{1/2}$~cm/s and $v_{\rm T} \sim \left(\alpha_{r} D_{\rm T}\right)^{1/2} \approx 1 \times \alpha_{r,-2}^{1/2}$~cm/s for the Hall and thermoplastic wavefronts, respectively. Here $c$ is the speed of light, $e$ is the elementary charge, $\kappa$ is the thermal conductivity, $C_{V}$ is the heat capacity, and $\alpha_{r}$ denotes the viscoelastic relaxation rate of the crust. The authors found in their simulations that in regions where the ambient field was closer to the critical value necessary for overstraining, thermoplastic waves become the prevailing mode of failure propagation. This feature is significant for stars with dominant toroidal fields (which we will argue is the case for E1547 in Sec. \ref{sec:ellip}), as these regions are more volatile.}

{With the above in mind, the scenario we envision for an E1547-like event proceeds as follows. A minor perturbation results in shallow fractures that are initially localised near the poles, warping the poloidal field and shutting off radio emission (see Sec. \ref{sec:shallow}). An excited Hall wave-chain then creeps towards the toroidally-dominated equator on a $\sim$month-long timescale -- plausible for rates $\alpha_{r} \gtrsim 10^{-2} \text{ s}^{-1}$ in regions of low electron number density $n_{e}$. Upon reaching this sector, where the field is stronger and near critical, the faster thermoplastic waves take over and the perturbation grows rapidly. Ignition chains in the high magnetic-energy-density region then trigger a glitch (Sec. \ref{sec:glitch}) and a visible X-ray burst (Sec. \ref{sec:globalquake}). A discussion of the applicability of the above to other systems and caveats more generally are provided in Sec. \ref{sec:discussion}. }

\section{Hydromagnetic structure}
\label{sec:hydrostruc}

Before presenting quantitative comparisons with the 2022 event data from E1547, we introduce an equilibrium model for the star. We work within a Newtonian framework and ignore rotational corrections for simplicity. The latter assumption is justified since E1547 -- despite being the second most rapidly-rotating Galactic magnetar known -- only spins at a rate of $\nus = 0.479\,$Hz \citep{low23}. In spherical coordinates, an axisymmetric magnetic field admits a Chandrasekhar decomposition of the form \cite[e.g.,][]{suvk19}
\begin{equation} \label{eq:magfield}
\bb = \Bs \left[ \nabla \psi \times \nabla \phi + \left( \frac{E_{\rm pol}}{E_{\rm tor}} \frac{1-\Lambda}{\Lambda} \right) ^{1/2} \beta(\psi) \nabla \phi \right],
\end{equation}
where $\psi$ is the magnetic streamfunction, $\beta$ defines the toroidal component of the field,  and $\Bs$ denotes the \emph{equatorial} field strength (i.e., half of the polar value). Typically, $B_{\star}$ is estimated through timing observations, as a surface field strength can be inferred from spin-down given a torque model. For example, one may deduce from the dipolar, plasma-filled magnetosphere model of \cite{spit06} that
\begin{equation} \label{eq:best}
B_{\star} \approx 3.2 \times 10^{19} \sqrt{- \frac{ I_{45} \dot{\nu}} {R_{6}^6 \left(1 + \sin^2 \alpha \right) \nus^3 } } \text{ G},
\end{equation}
for moment of inertia $I_{45}$ (in units of $10^{45} \text{g cm}^{2}$) and inclination angle, $\alpha$. Taking $\alpha = 160^{\circ}$, as suggested by \cite{cam08} who fitted the rotating vector model to polarisation data, and the (pre-outburst) spin-down value $\dot{\nu} = -3.63(3) \times 10^{-12} \text{ s}^{-2}$ \citep{low23}, estimate \eqref{eq:best} returns $B_{\star} \approx 1.5 \times 10^{14}$~G. The toroidal prefactor in \eqref{eq:magfield} controls the relative strength of the poloidal and toroidal components, with internal magnetic energies $E_{\rm pol}$ and $E_{\rm tor}$ respectively (see Appendix \ref{sec:ellipticity}), with $0 < \Lambda \leq 1$ setting their ratio: the star has a fraction $\Lambda$ of its magnetic energy concentrated in the poloidal sector, and $1- \Lambda$ in the toroidal sector. 

\begin{figure*}
\begin{center}
\includegraphics[width=\textwidth]{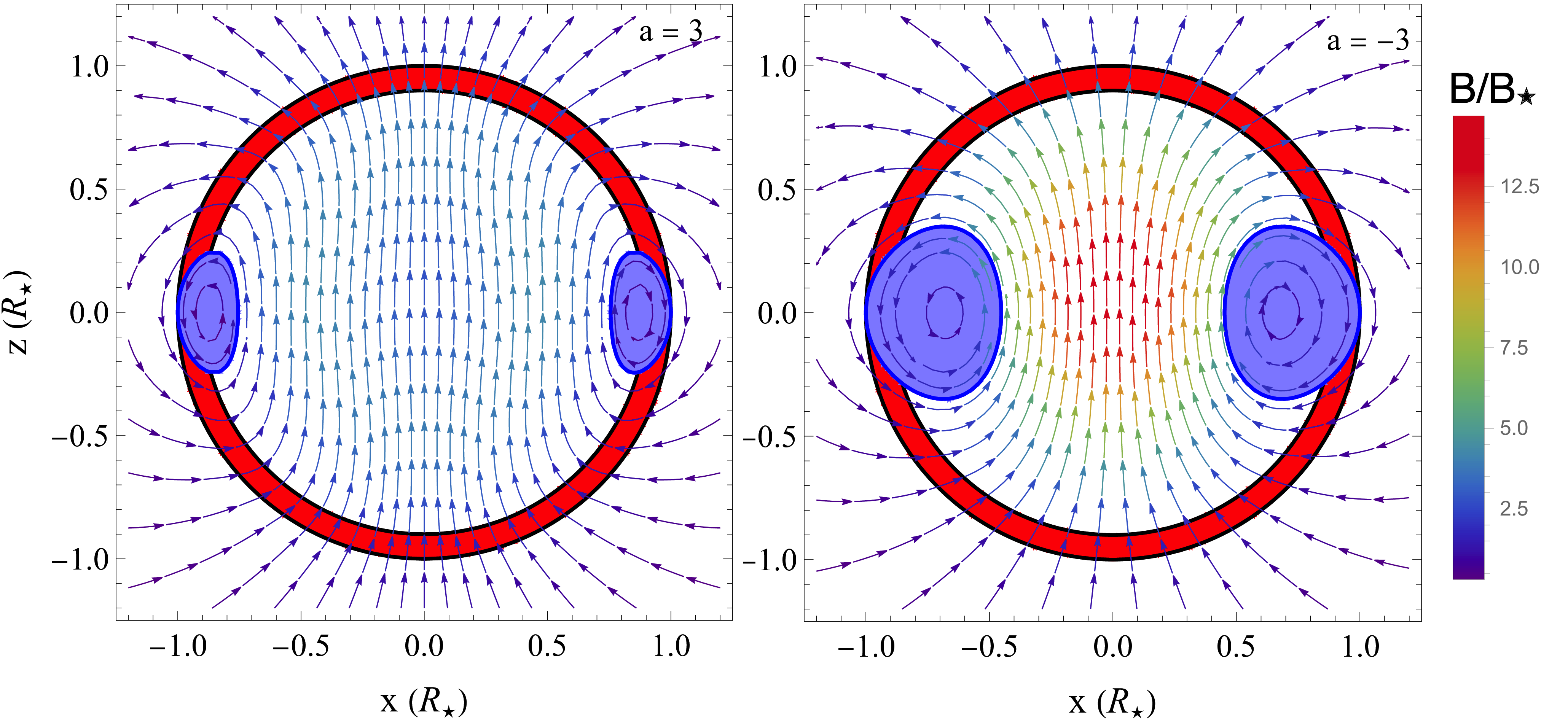}
\end{center}
\caption{Cross-sections of the magnetic field structure \eqref{eq:magfield} for two different configurations, defined through expression \eqref{eq:ffunc} with $a=3$ (left) and $a=-3$ (right). The toroidal field, if present, is confined to the blue-shaded ellipsoids at the equator. The crust, occupying the region $0.9 \Rs \leq r \leq \Rs$, is portrayed by the red annulus. The colour scale shows the relative strength of the \emph{poloidal} field ($\Lambda =1$), which is more core dominated for larger, negative values of $a$. The toroidal field migrates more into the crust for larger, positive $a$.}
\label{fig:magfields}
\end{figure*}

In general, one must specify an equation of state relating the hydrostatic pressure to the mass density, and then solve the energy and momentum balance (i.e., Grad-Shafranov) equations to determine the hydromagnetic structure of the star. However, given that the ratio of magnetic to gravitational-binding energies is $\ll 1$ for fields well below the virial limit $\sim 10^{18}$~G, it is reasonable to treat the magnetic field as a perturbation on a spherically symmetric background. Adopting this approximation, we follow the method described by \cite{mast11} to construct an equilibrium in a stratified, non-barotropic star. We refer the reader to the aforementioned reference for a thorough discussion on the validity of non-barotropic equilibria \cite[see also][]{mlm13,akg13,msm15}. Given a field \eqref{eq:magfield}, the magnetically-perturbed pressure, $\delta p$, and density, $\delta \rho$, are found through
\begin{equation} \label{eq:perteuler}
\nabla \delta p + \delta \rho \nabla \Phi = \frac{1}{4\pi} \left( \nabla \times \bb \right) \times \bb,
\end{equation}
where we have assumed the Cowling approximation ($\delta \Phi = 0$). 

Considering a dipole field for simplicity\footnote{Although we could apply the method of \cite{msm15} to self-consistently calculate deformations for a mixed poloidal-toroidal field of arbitrary multipolarity, the added complexity is unnecessary to illustrate the model put forth here.}, we write $\psi = f(r) \sin^2\theta$ for some radial function $f$. We choose the functions $f$ and $\beta$, following \cite{mast11}, by requiring that (i) the poloidal component of $\bb$ matches continuously to a force-free dipole at the stellar surface, (ii) the toroidal field is confined to the neutral curves of the poloidal field, and (iii) the current density, $\bj = \nabla \times \bb$, remains finite and continuous everywhere. Requirement (ii) is satisfied by picking $\beta = \Rs^{-3} \left(\psi-\psi_{c}\right)^2 \Theta\left(\psi/\psi_{c} - 1\right)$, where $\psi_{c}$ is defined by the $\psi$ value of the last poloidal field line closing inside the star. This choice is similar to others made in the literature \cite[e.g.,][]{cr13}. Even with these constraints there are infinitely many choices for $f$, reflecting the vast landscape of non-barotropic equilibria. {Indeed, as shown by \cite{gl16}, one is free to prescribe an arbitrary axisymmetric field when considering a normal fluid star that is non-barotropic. This is no longer true for non-axisymmetric or superfluid equilibria however (see their Table 1), and additional care must be taken in constructing an MHD model.}

%defined in terms of a normalised radial coordinate, $x = r/ \
In this work, we stipulate that the 2022 event in E1547 was triggered by magnetic reconfiguration(s). In reality, energy redistributions must be modelled by numerically solving the relevant time-dependent MHD equations, including elastic/plastic terms in the crust. We instead adopt a simple, analytic approach where the star `evolves' quasi-statically through a sequence of equilibria, as in the approach of \cite{lan15}. In particular, we consider a family of functions $f$ satisfying the aforementioned criteria, viz.
\begin{equation} \label{eq:ffunc}
\begin{aligned}
f_{a}(r) =& \Rs^2 \Bigg[ \left(\frac{35}{8}-a\right) \left(\frac{r}{\Rs}\right)^2  - \left(\frac{21}{4}-3a\right)\left(\frac{r}{\Rs}\right)^4 \\
&+ \left(\frac{15}{8} - 3a \right) \left(\frac{r}{\Rs}\right)^6 + a \left(\frac{r}{\Rs}\right)^8 \Bigg],
\end{aligned}
\end{equation}
which reduces to the \cite{mast11} choice for $a =0$. An equilibrium sequence is then defined through successive values of the parameters $a$, $B_{\star}$, and $\Lambda$, which control the geometry of the internal field. Cross sections of the field structure are shown in Figure \ref{fig:magfields} for $a = 3$ (left) and $a=-3$ (right).

\subsection{Quadrupolar ellipticity}
\label{sec:ellip}

Associated with the asphericity induced by the Lorentz force is a mass quadrupole moment, expressed through the \emph{ellipticity}, $\epsilon = \left(I_{zz} - I_{xx} \right) / I_{0}$, for moment of inertia (tensor) $I_{0}$ ($I_{ij}$) \cite[e.g.,][]{cr13}. Given a choice for the $\mathcal{O}(B^{0})$ equation of state, this ellipticity can be calculated after solving the perturbed Euler equation \eqref{eq:perteuler}. Details of the ellipticity calculation, for a star with a Tolman-VII mass density profile $\rho(r) = \rho_{c}\left[1-(r/\Rs)^2\right]$ with $\rho_{c} = 15\Ms /(8 \pi \Rs^3)$, are presented in Appendix \ref{sec:ellipticity}. We find a fit of the form
\begin{equation} \label{eq:ellip}
\epsilon \approx \epsilon_{0} \left( 1 + \epsilon_{1} a + \epsilon_{2} a^2\right) \left(1 - \frac {\lambda_{0} + \lambda_{1} a + \lambda_{2} a^2 }{\Lambda} \right) ,
\end{equation}
where $\epsilon_{0} = 2.27 \times 10^{-7} \times B_{\star,14}^2 M_{1.4}^{-2} R_{6}^{4}$, $\epsilon_{1} = -7.65 \times 10^{-2}$, $\epsilon_{2} = 1.18 \times 10^{-2}$, $\lambda_{0} = 0.351$, $\lambda_{1} = -1.30 \times 10^{-2}$, and $\lambda_{2} = -8.89 \times 10^{-4}$. Expression \eqref{eq:ellip} matches direct integrations to within $\sim 1\%$ at worst for $|a| \leq 3$. Note that setting $a =0$ returns the ellipticity of \cite{mlm13} [equation (7) therein]. Had we considered a superconducting core, $\epsilon_{0}$ would be boosted by a factor $\propto H_{c1}/B_{\star}$, where $H_{c1} \lesssim 10^{16}\,$G is the microscopic critical field \citep{lan13}.

%we could set $\epsilon_{0} \sim 10^{-7} \times B_{\star,12} H_{c1, 16} M_{1.4}^{-2} R_{6}^{4}$ to approximately recover the results 

\cite{mak16} interpreted phase modulations in the Suzaku X-ray data of E1547, with a periodicity of $T = 36^{+4.5}_{-2.5}\,$ks, as being evidence for free precession. Matching this to the expected free precession period, $T \gtrsim P_{\star}/|\epsilon|$ \citep{g70}, implies an ellipticity of magnitude $|\epsilon| \gtrsim 5.8^{+0.4}_{-0.7} \times 10^{-5}$. Using expression \eqref{eq:best} with the pre-outburst values (denoted with a subscript $i$) for E1547 quoted in Sec. \ref{sec:hydrostruc}, expression \eqref{eq:ellip} implies $\Lambda_{\rm i} \lesssim 0.01$ for any $|a| \leq 3$, i.e., the toroidal energy must comprise $\gtrsim\,99\%$ of the total magnetic energy (unless the core is superconducting). Such a partition is, however, permitted within MHD stability limits \citep{akg13}, and is consistent with the expectation that strong toroidal fields are necessary for instigating frequent magnetar activity \citep{pons11}, as is observed in E1547 \citep{coti18}. An internal field of strength $B_{\phi}\gtrsim10^{16}\,$G could also theoretically delay the onset of neutron superfluidity beyond $\sim700\,$yr, the characteristic age estimate for E1547 \citep{coti18}, thereby favouring a magnetic glitch mechanism (see Sec. \ref{sec:glitch}). We thus fix $\Lambda_{\rm i} = 0.01$. An important feature of expression \eqref{eq:magfield} is that the coefficient $E_{\rm pol}/ E_{\rm tor}$ itself depends on $a$ implicitly. As can be seen in Fig. \ref{fig:magfields}, the internal geometry of both sectors changes as a function of $a$, meaning that a given magnetic energy corresponds to different values of $\Lambda$ for varying $a$; see Appendix \ref{sec:ellipticity}.

Equipped with the expressions derived in this Section, we turn to the events surrounding E1547 in March/April 2022. 

\section{Outburst event in 1E 1547.0--5408}
\label{sec:outburst}

As detailed by \cite{low23}, radio pulsations from E1547 seemingly halted on March 16 2022, having otherwise been persistent since 2013. Three weeks following this cessation, short X-ray burst(s) were detected by the Fermi Gamma-Ray Burst monitor (GBM; April 3 2022) and the Swift Burst Alert Telescope (BAT; April 7 2022). The source was subsequently monitored by The Neutron star Interior Composition ExploreR (NICER). Although the precise time is difficult to ascertain owing to the absence of radio pulsations, a modest glitch ($\delta \nu \approx 0.2 \pm 0.1 \mu$Hz) coincided\footnote{{Short of timing data, one can only definitively say that the glitch occurred at some point during radio silence. However, \cite{low23} argue that the most probable onset time is indeed coincident with the X-ray burst on April 3. The magnitude of the glitch they recover depends sensitively on the onset date though, varying by up to an order of magnitude if one is agnostic (see their Section 3.2). While it is most natural in our model for a glitch to coincide with a burst (see Sec. \ref{sec:glitch}), there is a timescale involved in the conversion of magnetoelastic energy to X-rays in reality and a modest delay is tenable \citep{msm15b}.}} with the onset of bursting. The X-ray flux peaked at $F_{\rm X} \approx \left( 6.0 \pm 0.4 \right) \times 10^{-11} \text{erg s}^{-1} \text{cm}^{-2}$, decreasing steadily back down to a factor $\sim 3$ lower value, matching the persistent flux for this source since 2009, over $\approx$ 17 days. This allows us to estimate the energy contained in the overall burst, 
\begin{equation} \label{eq:bursten}
E_{\rm burst} \approx 4\pi d^2 \int dt \left(F_{\rm X, burst} - F_{\rm X, persistent}\right),
\end{equation}
for source distance $d$, a quantity which is uncertain; expansion rates from the dust-scattering rings following the 2009 outburst suggest $4 \lesssim d/\text{kpc} \lesssim 5$ \citep{tiengo10}, though updated dispersion measure measurements point towards $5.9 \lesssim d/\text{kpc} \lesssim 8.3$ \citep{low23}. We thus have $1 \lesssim E_{\rm burst}/\left(10^{41}\,\text{erg}\right) \lesssim 6$. By contrast, the burst energy of the 2016 event from PSR J1119--6127 was about an order-of-magnitude lower, $E_{\rm burst} \approx 5 \times 10^{39}\,$erg \citep{gog16}.

 Two weeks after the X-ray flux stabilised to its persistent (post 2009) value the source recovered as a radio pulsar, though was spinning down at a steeper rate. In March 2022 the source displayed $\dot{\nu} = -3.63(3) \times 10^{-12} \text{ s}^{-2}$, though this jumped to $\dot{\nu} + \delta \dot{\nu} \approx -6 \times 10^{-12} \text{ s}^{-2}$ just after the burst, followed by a linear decrease \cite[though in a noisy fashion that is typical for magnetars;][]{mel99} over the following $\gtrsim 100$ days to reach $\approx -8 \times 10^{-12} \text{ s}^{-2}$. 

\subsection{Shallow fractures: radio switch off}
\label{sec:shallow}

The first phase of the 2022 event in E1547 concerns the initial radio switch-off. As noted by \cite{low23}, this phenomena could point towards an (undetected) slow rise in high-energy activity in the intervening weeks prior to the April burst(s). {We suggest here that a} relatively mild perturbation occurred in $\sim\,$March 2022, which only led to shallow, localised fracturing in polar regions. If fractures are localised to the polar caps, the energy released may not be large enough to cause a noticeable increase in X-ray activity, which is relatively high for E1547 \citep{coti18}. Magnetospheric gap zones may be congested by crustal ejecta or fireball plasma though, preventing the requisite pair-production multiplicity from being met to permit coherent radio emission \cite[assuming a polar-gap model;][]{rs75}. This and similar interpretations have been put forth for the radio shutdown in the high-$B$ pulsar PSR J1119--6127 following its outburst in 2016 \citep{arch17}.

Whether (and where) a crust breaks given a magnetic reconfiguration is a complicated problem that has been studied by several authors \cite[e.g.,][]{lan15,suvk19,koj22}. The response of the crust to stress depends primarily on the crustal shear modulus, $\mu$, and the eigenfunction of the Lagrangian perturbation vector. Assuming that the perturbation is sourced by a magnetic shift, $\bb \rightarrow \bb + \bdb$, the square of the Cauchy stress tensor can be written as \citep{lan15} \cite[though cf.][who use a slightly different form]{lg19}
\begin{equation} \label{eq:stresstensor}
\sigma_{ij}\sigma^{ij} = \frac{2 \delta B^2 B^2 +3 B^4 +3 \delta B^4 - 8 \left( \delta \boldsymbol{B} \cdot \boldsymbol{B} \right)^{2} }{64 \pi^2 \mu^2} .
\end{equation}
If the von Mises criterion applies to a neutron star crust \cite[cf.][]{baiko18}, faults occur in places such that
\begin{equation} \label{eq:vonmises}
\sigma = \sqrt{\frac{\sigma_{ij} \sigma^{ij}}{2}} \geq \sigma_{\rm max},
\end{equation}
where $\sigma_{\rm max}$ is an elastic maximum set by the crustal microphysics. For a polycrystalline crust, \cite{baiko18} estimate that $\sigma_{\rm max} \approx 0.04$. We adopt this value here.

Suppose now that the star undergoes a quasi-static evolution from one set of values $\{B_{\star,\rm i}, a_{\rm i}, \Lambda_{\rm i}\}$ to another $\{B_{\star,\rm f}, a_{\rm f}, \Lambda_{\rm f}\}$. This translates into a strain through expression \eqref{eq:stresstensor}, from which we can determine if and where the crust yields via the von Mises criterion \eqref{eq:vonmises}. In general, we require that a quasi-evolution $\{X_{\rm i}\} \rightarrow \{X_{\rm f}\}$, occuring within $\sim\,$weeks, is such that (i) the global magnetic energy is conserved (within a numerical tolerance of $< 0.01\%$), and (ii) the energy subsequently released in fracture zones is positive (see below).

A particular realisation is shown in Figure \ref{fig:firstquake}, using the shear modulus profile of \cite{lan15} described in Appendix \ref{sec:ellipticity}. The colour scale shows the strain magnitude $\sigma$, defined in \eqref{eq:vonmises}, where we pick $B_{\star,\rm f} = 0.999 B_{\star,\rm i}$, $a_{\rm i} = 0$, $\Lambda_{\rm i} = \Lambda_{\rm f} = 0.01$, and $a_{\rm f} = -0.01$. In particular, once radio pulsations ceased and timing data became unavailable, it is plausible that the surface field grew either marginally stronger (faster spin-down) or weaker (slower spin-down; see also Sec. \ref{sec:acceleratedsd}). Quantitatively similar results are obtained for a marginally larger $B_{\star, \rm f}$ instead with a small, positive $a_{\rm f}$. The shaded zones near the poles $(r \approx \Rs, \theta \approx 0,\pi)$ depict regions where inequality \eqref{eq:vonmises} is satisfied (where we have $\sigma \approx 0.05$), indicating a crustal failure. Note the deep stress near the equator $(r \gtrsim 0.9 \Rs, \theta \sim \pi/2)$ arises from the intense toroidal field (as we set $\Lambda = 0.01$), though $\sigma < \sigma_{\rm max}$ there.

For the case depicted in Fig. \ref{fig:firstquake}, the fracture zones at the poles comprise $2.6\%$ of the crustal volume. We estimate the magnetic energy release through \citep{lan15}
\begin{equation} \label{eq:magenergyrelease}
E_{\rm quake} = \int_{\sigma \geq \sigma_{\rm max}}  dV \frac{\left(B^2 - \delta B^2\right)} {8 \pi},
\end{equation}
finding $E_{\rm quake} = 2.4 \times 10^{40} \text{erg}$. Accounting for an efficiency $\eta$ for converting magnetic energy into X-rays, we have $E_{\rm quake}/E_{\rm burst} \sim 10^{-3} \times (\eta/0.01)$; likely invisible relative to noise (though note that X-ray flux data are unavailable in the days between radio shut-off and the burst; see Figure 2 in \citealt{low23}). Had we taken $B_{\star,\rm f} = 0.9995 B_{\star,\rm i}$ and a marginally greater $a_{\rm f}$ instead (to ensure energy conservation), the fractures are still confined to the poles (radio shut-off), though we find an even lower value of $E_{\rm quake} = 1.4 \times 10^{39} \text{erg}$. %These values for $E_{\rm quake}$ are comparable to the ; if $\Lambda \lesssim 10^{-2}$ for this star also, a scenario like that depicted in Fig. \ref{fig:firstquake} could apply to that event.

\begin{figure}
\begin{center}
\includegraphics[width=0.497\textwidth]{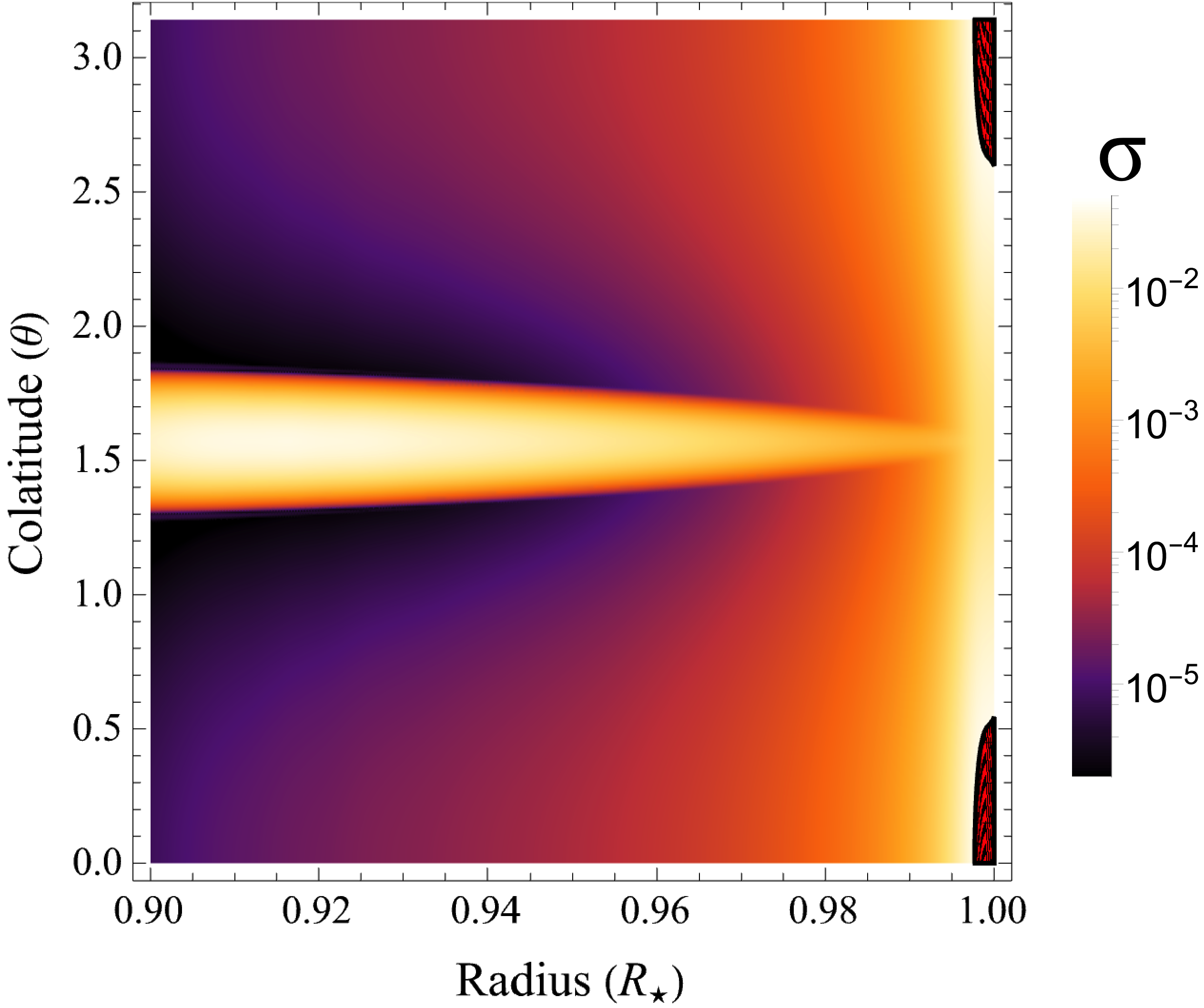}
\end{center}
\caption{Crustal strain induced by a mild perturbation to the poloidal magnetic field, as detailed in the main text. The shaded regions near the polar caps ($r \approx \Rs, \theta \approx 0,\pi$) represent zones where the von Mises criterion \eqref{eq:vonmises} is satisfied, indicating a crustal failure. Crustal ejecta from these zones, or pair-plasma from a fireball, may pollute the polar gaps, turning the pulsar off for a time.}
\label{fig:firstquake}
\end{figure}

\subsection{Glitch: quadrupolar shifts}
\label{sec:glitch}
%built up from the seed perturbation
{The minor event depicted in Fig. \ref{fig:firstquake} could serve as a precursor to a more dramatic reconfiguration, taking place $\sim 3$~weeks later, via the mechanism described in Sec. \ref{sec:hall}. Such a reconfiguration, and magnetic energy releases more generally, would adjust the stellar ellipticity \eqref{eq:ellip} since $\epsilon \propto B^2$.} As described by \cite{i01,grs15}, a rapid change in the magnetically-induced ellipticity might be responsible for a sudden spin-frequency jump: a shift $\epsilon_{\rm i} \rightarrow \epsilon_{\rm f}$ induces a change in $\nus$ of order
\begin{equation} \label{eq:glitchfreq}
\frac{\delta \nu}{\nus} \approx \frac{2}{3} \left( \epsilon_{\rm i} - \epsilon_{\rm f} \right),
\end{equation}
by angular momentum conservation. A positive glitch implies $\epsilon_{\rm i} > \epsilon_{\rm f}$, suggesting that poloidal decay (or rearrangement; it is not strictly necessary that $E_{\rm pol,i} > E_{\rm pol,f}$) was primarily responsible for powering the burst since poloidal energy contributes positively to $\epsilon$. Although the model of \cite{grs15} was initially put forth as an explanation for an \emph{anti-glitch} seen in 1E 2259+586 \cite[see also][]{i01,msm15b}, it also applies as a model for magnetar glitches since positive or negative shifts in the stellar oblateness are plausible from a field reconfiguration {triggered by an instability in a young object, rather than gradual Ohmic losses (see Sec. \ref{sec:discussion})}. %\citep{hm15}.

Using expression \eqref{eq:ellip}, we can invert \eqref{eq:glitchfreq} to infer the post-glitch magnetic parameters as a function of the initial ones and the magnitude of $\delta \nu$. In particular, this allows us to fix $\Lambda_{\rm f}(a_{\rm i},a_{\rm f},B_{\star, \rm i},B_{\star, \rm f},\Lambda_{\rm i}, \delta \nu)$, though the resulting expression is lengthy so we avoid writing it out explicitly. Given the estimate on the ellipticity of the source from precession provided by \cite{mak16}, $|\epsilon| \approx 6 \times 10^{-5}$, {changes only on the order of $\sim 1\%$} in the {quadrupole moment} are necessary to produce a glitch of order $\delta \nu / \nu_{\star} \approx \left(4 \pm 2 \right) \times 10^{-7}$.

\subsection{Accelerated spin-down}
\label{sec:acceleratedsd}

The field strength, $B_{\star, \rm f}$, {could theoretically} be fixed if we attribute the accelerated spin-down following the glitch to a surface field increase; a reconfiguration need not preserve the field structure at $r\geq\Rs$. Using expression \eqref{eq:magfield} but with $\dot{\nu} \rightarrow \dot{\nu} + \delta \dot{\nu} \approx -6 \times 10^{-12}\,\text{s}^{-2}$ and $\nus \rightarrow \nus + \delta \nu$, we find the larger value $B_{\star, \rm f} \approx 2 \times 10^{14}\,$G. Note, however, that such an interpretation is almost certainly oversimplified: twists imparted by crustal failures work to inflate magnetospheric field lines, thereby increasing the Poynting flux and accelerating spin-down \citep{thom02,belo09}. Ejected plasmoids can also comb open previously closed lines. These effects likely explain the source's accelerated spin-down, especially given the continued decrease in $\dot{\nu}$ after the glitch; twist-induced torque enhancements can take $\gtrsim 10^2\,$days to peak \citep{parfey13}. {Although a particular function $B_{\star}(t)$ could be engineered such that the $\dot{\nu}(t)$ and $\ddot{\nu}(t)$ profiles observed by \cite{low23} are reproduced for a given torque model, the resulting curve would not be physically motivated. We therefore do not model the accelerated spin-down here, though note that this feature fits naturally within the crustal failure picture \citep{belo09}.} %Intuitively, the high electrical conductivity of the magnetosphere implies that if plasmoids are ejected outward, previously closed lines may be combed open, in turn leading to greater angular momentum losses. 

\subsection{Deeper fractures: X-ray flare}
\label{sec:globalquake}

Assuming the quake and glitch {occurred simultaneously (cf. Footnote 2)}, we are now tasked with finding values of $a_{\rm i}$ and $a_{\rm f}$ such that the ellipticity changes by the amount demanded by the glitch \eqref{eq:glitchfreq}, while ensuring that the global magnetic energy is conserved and that the quake energy \eqref{eq:magenergyrelease} is positive. This is achieved by performing a scan over values $|a| \leq 3$.

\begin{figure}
\begin{center}
\includegraphics[width=0.497\textwidth]{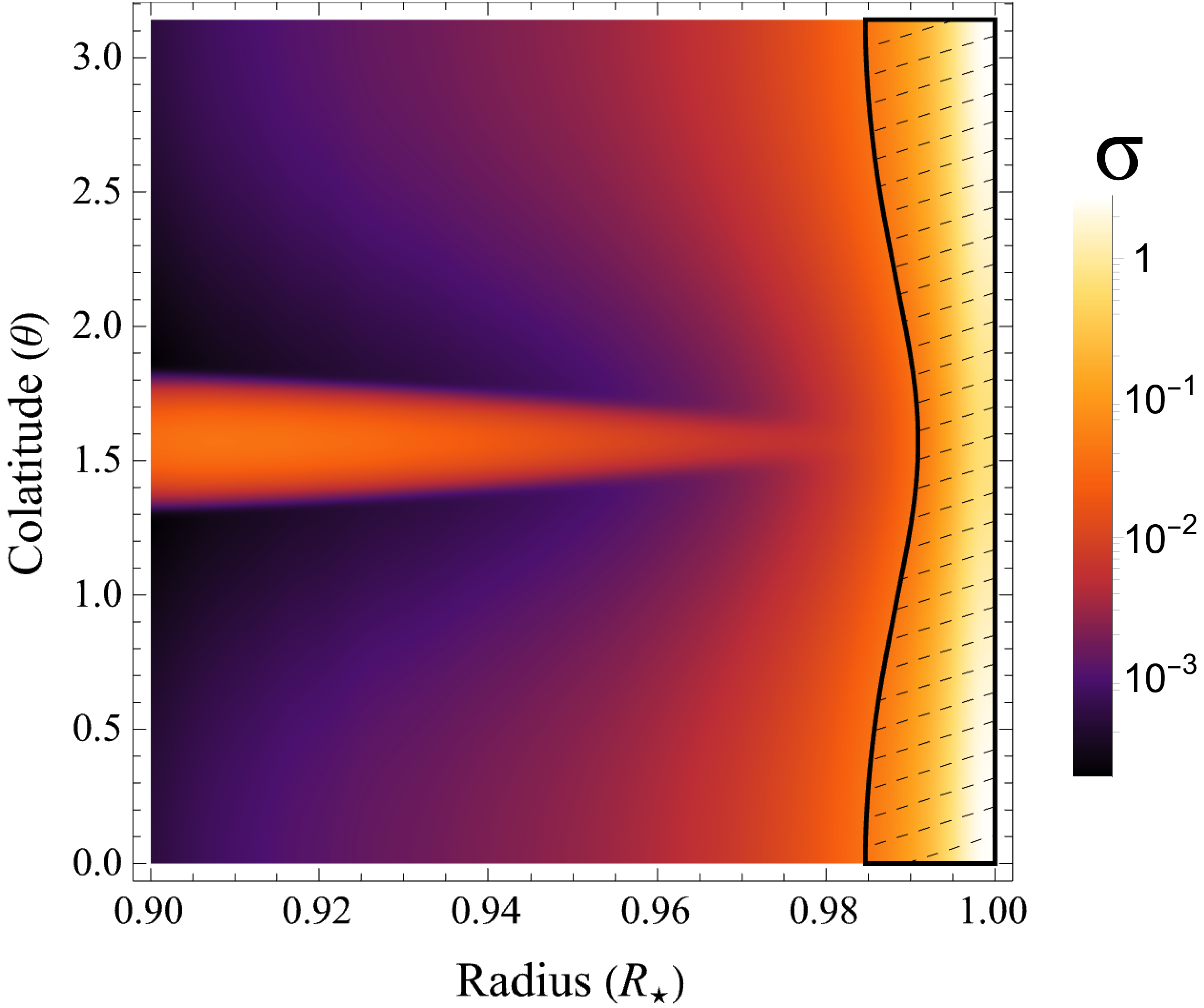}
\end{center}
\caption{Similar to Fig. \ref{fig:firstquake}, though depicting a more intense reconfiguration where the entire layer $r \gtrsim 0.98\Rs$ exceeds the von Mises threshold. Magnetic parameters are chosen such that the quadrupolar ellipticity $\epsilon$ changes by the amount predicted by the glitch (equation \ref{eq:glitchfreq}), with $\Lambda_{\rm i} = 0.01$ chosen to approximately match the precession-implied ellipticity of \protect\cite{mak16}.}
\label{fig:largequakenotor}
\end{figure}

{The resulting fracture pattern for one such combination, with $a_{\rm i} = -3$, $a_{\rm f} = -2.82$, and $B_{\star,\rm f} = 1.43 \times 10^{14}$~G (i.e., assuming accelerated spin-down is due to magnetospheric torques), is depicted in Figure \ref{fig:largequakenotor}. Expression \eqref{eq:glitchfreq} implies that we require $\Lambda_{\rm f} \approx 0.008$ to match the (lower limit of the) glitch data. Because the toroidal volume shrinks ($a_{\rm f} > a_{\rm i}$; see Fig. \ref{fig:magfields}), the change in toroidal energy is $\approx 2\%$. The entire outer layers of the crust ($r \gtrsim 0.98 \Rs$) succumb to Maxwell stresses, with a shallower valley around the equator ($\theta \sim \pi/2$). One eighth of the crustal volume exceeds the von Mises limit \eqref{eq:vonmises} in this case, releasing a sizeable quake of energy $E_{\rm quake} = 4.1 \times 10^{43} \text{ erg}$. For an X-ray conversion efficiency of $\eta \sim 1\%$, this matches the E1547 burst energy \eqref{eq:bursten}.} Given the lack of hard X-ray evolution or low-frequency quasi-periodic oscillations (i.e., torsional modes) following the burst however -- often observed in global-scale magnetar outbursts \citep{thom17}  -- even these parameter choices may be too extreme. {It is relatively straightforward to choose parameters that lead to phenomenologically similar but milder quakes. For instance, a larger stress threshold of $\sigma_{\rm max} \sim 0.1$, as predicted by the molecular dynamics simulations of \cite{hk09}, reduces the quake energy by a factor $\gtrsim 2$ (see also Sec. \ref{sec:discussion}).} Greater disparities $|a_{\rm i} - a_{\rm f}|$  and/or $|B_{\star, \rm i} - B_{\star, \rm f}|$ predict similar results but with larger glitches. {Considerably larger changes in toroidal strength, $|\Lambda_{\rm i} - \Lambda_{\rm f}|$, cause deep gorges to fail around the equator, leading to giant quakes commensurate with the energy output of the 2004 hyperflare from SGR 1806--20 \cite[$\sim 2 \times 10^{46} \text{ erg}$;][]{palm05}.}

Once the magnetosphere recovers to a quasi-steady state the object may then reactivate as a radio pulsar, similar to the case of the 2016 burst(s) in PSR J1119--6127 \citep{arch17}.  

\section{Discussion}
\label{sec:discussion}

In this short paper, we have offered an interpretation for the 2022 outburst event from E1547 reported by \cite{low23} in terms of a ramping magnetic reconfiguration. An initial perturbation inside the star leads to critical Maxwell stresses being applied just beneath the polar caps. The shallow fractures that form could spark a pair-plasma fireball, particles from which come to pollute the acceleration gaps, extinguishing radio activity (Sec. \ref{sec:shallow}). Such a picture is similar to that proposed by \cite{arch17} for the 2016 event from PSR J1119--6127. Importantly, though, these fractures may be shallow enough in the case of E1547 that they do not reveal themselves against the high baseline X-ray flux from the source. Stress continues to build up \cite[perhaps via Hall-wave avalanches {as described in Sec. \ref{sec:hall}};][]{xi16}, culminating in a global event where magnetic energy is rapidly redistributed such that the stellar moment of inertia shifts slightly, inducing a glitch (Sec. \ref{sec:glitch}) and accelerating spin-down (Sec. \ref{sec:acceleratedsd}), while fuelling the observed X-ray burst (Sec. \ref{sec:globalquake}). The source then returns to a radio-loud state once the magnetosphere recovers.

At least within the context of the simple, quasi-static models presented here, Figs. \ref{fig:firstquake} and \ref{fig:largequakenotor} demonstrate that a wide variety of energetics and fracture geometries can be accounted for given a glitch size $\delta \nu/ \nus$. An even more diverse range could be realised if we relax some of the assumptions used here. (1) The $\sim36\,$ks modulations in Suzaku X-ray data need not necessarily point towards free precession \citep{mak16}, so that choices $\Lambda_{\rm i} \gtrsim 0.01$ could be permitted, substantially reducing magnetic energy releases. (2) Relaxing the force-free exterior condition, thereby allowing for magnetospheric currents to assist with accelerated spin-down, leads to different boundary conditions applied to the streamfunction $\psi$, and thus more freedom in setting the interior geometry \cite[see, e.g.,][]{glamp14}. (3) The von Mises maximum \eqref{eq:vonmises} may be considerably larger than 0.04 -- \citealt{hk09} estimate $\sigma_{\rm max} \sim 0.1$ from molecular dynamics simulations -- or even anisotropic, and vary from neutron star to neutron star. (4) The magnetic field need not be a simple dipole as we have considered (see Footnote 1), again enriching the configuration space. Observations of pulse profile evolution \citep{thom02} and spectral features seen in X-rays \citep{gog11} suggest that magnetar fields are comprised of non-negligible multipoles, as do crustal Hall-plastic-Ohm simulations \citep{lg19}. {Quadrupolar toroidal regions may also reside closer to the pole (see Figure 2 in \citealt{msm15}), softening requirements on the viscoelastic relaxation rate $\alpha_{r}$ (see below).} (5) Assuming a superconducting core boosts the ellipticity {by a factor $\lesssim 10^{2}$} \citep{lan13}, making it easier to produce large glitches.

Moreover, the intense stresses depicted in Fig. \ref{fig:largequakenotor} cannot be built up instantaneously in reality, as failed sections of the crust will flow plastically once even a small zone exceeds the stress threshold \citep{bell14}. As such, our results do not necessarily represent realistic astrophysical pre/post-dictions, but rather illustrate the plausibility of explaining the 2022 event observed in E1547 via a growing magnetic perturbation. In order to verify the scenario presented here, it would be necessary to build a simulation where the magnetar field is evolved over $\sim$month-long timescales, while accounting for dynamics and backreactions related to fracturing as {described in Sec. \ref{sec:hall}}: the Stokes-like plastic flow that the crust enters into after a failure can significantly alter the subsequent evolution \citep{lg19}. We leave this ambitious task to future work.

{How does the model suggested here fit in the context of other systems? A burst could instead precede radio activation (as seen, e.g., in XTE J1810--197) if a failure chain favourably twists the field lines near the dipole axis to allow $e^{\pm}$ cascading, as argued by \cite{belo09}. Furthermore, a glitch of either sign could accompany an outburst depending on whether poloidal or toroidal energy is relieved [equation \eqref{eq:glitchfreq}], both of which are plausible via MHD instabilities \citep{zink11} or magnetoplastic evolution \citep{gl21} in a young system like E1547, given the highly uncertain birth topology of magnetars. An observable burst might not even occur in bright systems if the energy release is sufficiently small so as to be invisible relative to noise, meaning the internal field evolves but $\sigma < \sigma_{\rm max}$ almost everywhere (which might explain the 2020 anti-glitch and radio emissions in SGR 1935+2154; \citealt{youn23}). By contrast, X-ray flaring without a glitch of any kind would be difficult to explain since large internal magnetic energy releases imply a shifting mass quadrupole moment [equation \eqref{eq:ellip}]. In this case an external mechanism, like that described by \cite{link14}, likely applies instead. The scenario put forth here for E1547 would also be difficult to accept if the relaxation rate $\alpha_{r}$ was much lower than assumed in Sec. \ref{sec:hall}, since disturbances could not reach the toroidal region(s) in time to explain the sequence of events. Quake geometries could also be tied to the inferred shape of newly-formed/broadened hotspots, in principle, though such modelling lies beyond the scope of this article. Overall however, in the absence of a gravitational-wave detection or some other means to probe the internal field directly, a wide variety of event sequences and energetics are plausible.}

A consequence of the model presented here, and the observations made by \cite{low23} more generally, is that small-scale fractures localised at the polar caps (or elsewhere) could be responsible for nulling in radio-loud magnetars. It may not be necessary that global events take place in order to suppress radio activity. If, for example, the magnetic field was so intense ($\lesssim 10^{16}\,$G) that the polar gaps are habitually contaminated, the pulsar may only rarely turn on, appearing as something akin to a rotating radio transient \cite[RRAT;][]{mc06}. Importantly, the magnetic energy released in this case may not be large enough to announce itself via an enhanced X-ray flux. This might help to explain the nature of the mysterious radio-loud object GLEAM-X J162759.5--523504.3, which appears to possess an ultra-strong magnetic field and yet is undetectable in X-rays \citep{hw22,suvm23}.

%%%%%%%%%%%%%%%%%%%%%%%%%%%%%%%%%%%%%%%%%%%
\section*{Acknowledgements}
This work was supported by the Alexander von Humboldt foundation. {I am grateful to the anonymous referee for feedback on matters of style and substance.}%I thank Marcus Lower for correspondence.

%%%%%%%%%%%%%%%%%%%%%%%%%%%%%%%%%%%%%%%%%%%

\section*{Data availability statement}
Observational data used in this paper are quoted from the cited works. No new data were produced here.

%%%%%%%%%%%%%%%%%%%%%%%%%%%%%%%%%%%%%%%%%%%%%%%%%%%%%%%%%%%%%%%%
%%%%%%%%%%%%%%%%%%%%%%%%%%%%%%%%%%%%%%%%%%%%%%%%%%%%%%%%%%%%%%%%

\bibliographystyle{mn2e}

%%%%%%%%%%%%%%%%%% APPENDIX %%%%%%%%%%%%%%%%%%%%%%%%%%%%%%%%%%%%%%%

\appendix
\section{Calculation of the ellipticity and crustal strain}
\label{sec:ellipticity}

This Appendix details the calculation of the ellipticity \eqref{eq:ellip} and the elastic stress \eqref{eq:stresstensor}, requiring a shear modulus as input.

%\subsection{Fitting procedure}
%The resulting fit matches the numerical data to within $\sim 1\%$ at worst.

Given a background hydrostatic model, the perturbed density in the equation of motion \eqref{eq:perteuler} can be found directly by taking the curl, yielding \citep{mlm13}
\begin{equation} \label{eq:perturbeddens}
\frac{\partial \delta \rho}{\partial \theta} = -\frac{r}{4 \pi \Rs} \left( \frac{d \Phi}{dr} \right)^{-1} \left\{ \nabla \times \left[ \left( \nabla \times \bb \right) \times \bb \right] \right\}_{\phi}.
\end{equation}
Integrating \eqref{eq:perturbeddens} up to an arbitrary function of radius allows us to compute the ellipticity, which reduces to \citep{mast11,msm15}
\begin{equation} \label{eq:ellipintegral}
\epsilon = \frac{\pi}{I_{0}} \int_{V} dr d \theta \delta \rho(r,\theta) r^4 \sin \theta \left( 1 -3 \cos^2\theta \right).
\end{equation}
For the configuration defined by expressions \eqref{eq:magfield} and \eqref{eq:ffunc}, the integral \eqref{eq:ellipintegral} cannot be evaluated analytically except in the case $\Lambda = 1$ (i.e., no toroidal field). The potential $\Phi(r)$ is found through the Poisson equation. In order to arrive at expression \eqref{eq:ellip} used in the main text, we evaluated \eqref{eq:ellipintegral} for $-3 \leq a \leq 3$ in steps of $\delta a = 0.01$ and performed a least squares fit on a function that accounts for terms up to $\mathcal{O}(a^2)$. Including cubic and higher-order terms reduces the maximum error below the $\lesssim1\%$-level quoted in the main text, though such accuracy is unnecessary for the simple model considered here. A similar procedure was carried out for the magnetic energy ratio, needed to infer the toroidal prefactor in equation \eqref{eq:magfield}. The poloidal and toroidal energies are defined as
\begin{equation}
E_{\rm pol} = \frac{1}{8\pi}\int_{V}\frac{dV}{r^2 \sin^2\theta} \left[ \left( \frac{1}{r} \frac{\partial \psi}{\partial \theta}\right)^2 + \left( \frac{\partial \psi}{\partial r} \right)^2 \right],
\end{equation}
and
\begin{equation}
E_{\rm tor} = \frac{1}{8\pi}\int_{V} dV \frac{ \beta(\psi)^2}{r^2\sin^2\theta},
\end{equation}
respectively. Fitting numerical integrations, we find
\begin{equation} \label{eq:energyratio}
\frac{E_{\rm tor}}{E_{\rm pol}} \approx \left(u_{0} + u_{1} a + u_{2} a^2 + u_{4} a^4 \right) e^{k a},
\end{equation}
with $u_{0} = 2.30 \times 10^{-6}$, $u_{1} = -5.20 \times 10^{-7}$, $u_{2} = 2.18 \times 10^{-7}$, $u_{4} = -9.71 \times 10^{-9}$, and $k = -0.774$. Expression \eqref{eq:energyratio} matches the numerical integrals to within $\sim 5\%$ at worst for $|a| \leq 3$.

\subsection{Shear modulus}

\begin{figure}
\begin{center}
\includegraphics[width=0.497\textwidth]{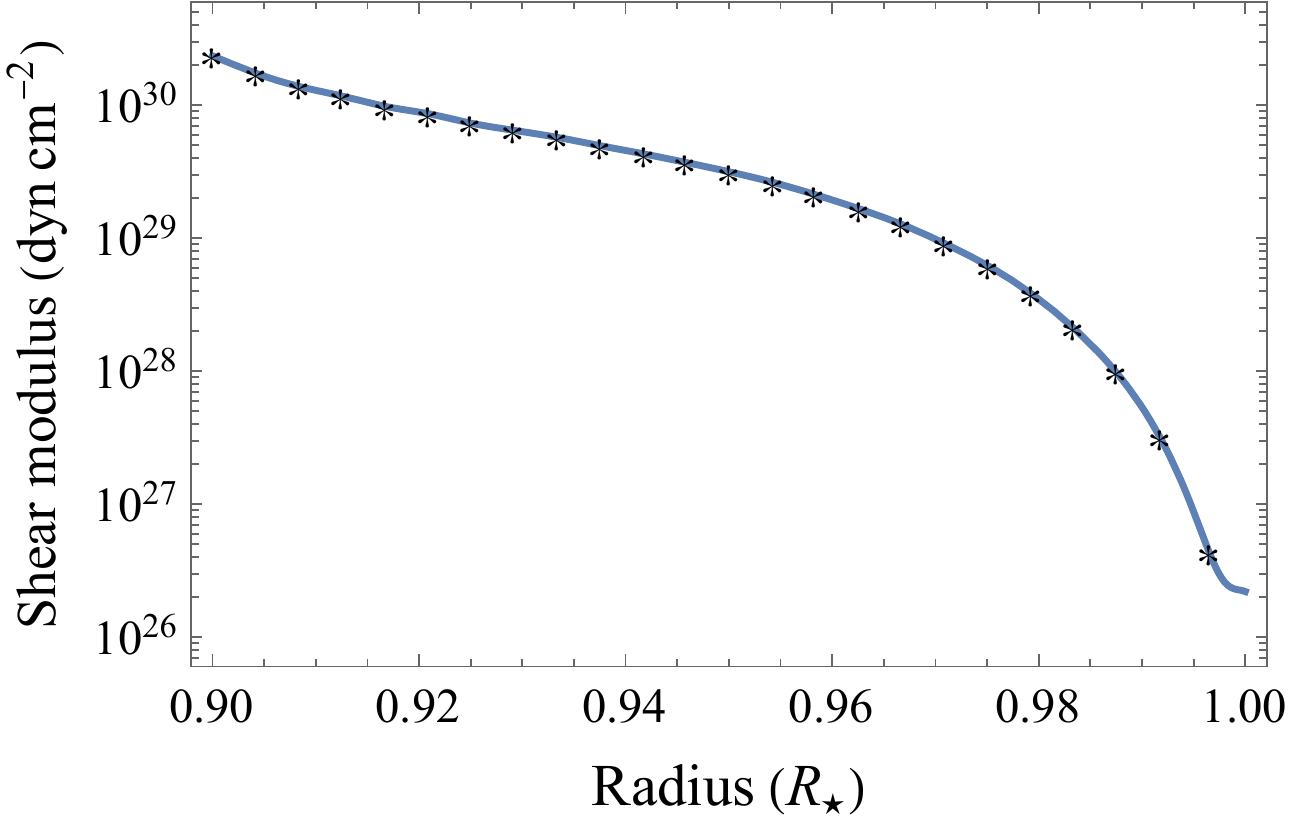}
\end{center}
\caption{Shear modulus, $\mu$, used in the main text to calculate the strain $\sigma_{ij}$ in a magnetar crust. Data points (black stars) are from \protect\cite{lan15}, based on references cited therein.}
\label{fig:shearmod}
\end{figure}

We estimate the shear modulus in a magnetar crust following \cite{lan15}, who fit a curve to molecular-dynamics simulation data, supplemented with a liquid drop equation of state, given in references cited therein. We use the same curve here, depicted in Figure \ref{fig:shearmod}. Note, however, that the value of the shear modulus in the very outer layers of the crust $(r \approx \Rs)$ has been extrapolated from the existing fit, {ignoring the possibility of a crustal ocean.}

%plot of shear modulus?
%%%%%%%%%%%%%%%%%%%%%%%%%%%%%%%%%%%%%%%%%%%%%%%%%%%%%%%%%%%%%

%%%%%%%%%%%%%%%%%%

\label{lastpage}

\end{document}